\documentclass[twocolumn]{aastex62}

\usepackage{apjfonts}

\shorttitle{Dwarf Barium-Star Nucleus of K 1-9}
\shortauthors{Bond et al.}

\newcommand{\BaII}{\ion{Ba}{2}}

\newcommand{\Ha}{H$\alpha$}
\newcommand{\Hb}{H$\beta$}

\newcommand{\OIII}{\ion{O}{3}}
\newcommand{\SrII}{\ion{Sr}{2}}

\newcommand{\sprocess}{{\it s}-process}

\def\baii{\ion{Ba}{2}}
\def\BaII{\ion{Ba}{2}}

\def\caii{\ion{Ca}{2}}

\def\hii{\ion{H}{2}}
\def\nii{\ion{N}{2}}

\def\oiii{\ion{O}{3}}

\def\srii{\ion{Sr}{2}}
\def\baii{\ion{Ba}{2}}

\def\Gaia{{\it Gaia}}
\newcommand{\GALEX}{{\it GALEX}}

\newcommand{\HST}{{\it HST}}

\newcommand{\TESS}{{\it TESS}}

\def\K{$\text{K\,1-9}$}
\def\Hen{$\text{Hen\,2-39}$}

\begin{document}

\title{Spectroscopic Survey of Faint Planetary-Nebula Nuclei. VIII\null. The Dwarf Barium Central Star of Kohoutek 1-9\footnote{Based on observations obtained with the Hobby-Eberly Telescope (HET), which is a joint project of the University of Texas at Austin, the Pennsylvania State University, Ludwig-Maximillians-Universit\"at M\"unchen, and Georg-August Universit\"at G\"ottingen. The HET is named in honor of its principal benefactors, William P. Hobby and Robert E. Eberly.} }

\author[0000-0003-1377-7145]{Howard E. Bond}
\affil{Department of Astronomy \& Astrophysics, Pennsylvania State University, University Park, PA 16802, USA}
\affil{Space Telescope Science Institute, 
3700 San Martin Dr.,
Baltimore, MD 21218, USA}

%\author[0000-0003-2071-2956]{Soumyadeep Bhattacharjee}
%\affiliation{Department of Astronomy, California Institute of Technology, 1216 E. California Blvd, Pasadena, CA, 91125, USA}

\author[0009-0005-3715-4374]{Peter Goodhew}
\affil{Deep Space Imaging Network, 108 Sutton Court Rd., London, W4 3EQ, UK}

%\author[0000-0002-9018-9606]{Dana Patchick}
%\affil{Deep Sky Hunters Consortium, 1942 Butler Ave. Los Angeles, CA 90025, USA }

\author[0009-0005-9964-1602]{Daniel Stern}
\affil{MEA Observatory,
16252 Andalucia Ln., 
Delray Beach, FL  33446, USA }

\author[0009-0009-3986-4336]{Jonathan Talbot}
\affil{Stark Bayou Observatory, 1013 Conely Cir., Ocean Springs, MS 39564, USA}

%\author[0000-0002-4964-4144]{John R. Thorstensen}
%\affil{Department of Physics \& Astronomy, 6127 Wilder Laboratory, Dartmouth College, Hanover, NH 03755, USA}

\author[0000-0003-2307-0629]{Gregory R. Zeimann}
\affil{Hobby-Eberly Telescope, University of Texas at Austin, Austin, TX 78712, USA}

\correspondingauthor{Howard E. Bond}
\email{heb11@psu.edu}

\begin{abstract}

In the course of our ongoing survey of faint planetary-nebula nuclei (PNNi), we obtained optical spectroscopy of the central star of the little-studied PN Kohoutek~1-9 (\K). Its spectrum is found to be that of a G-type dwarf with strong absorption features of carbon molecules and \sprocess\ elements such as Sr and Ba---a dwarf barium star. \K\ thus joins a very small group of PNe with barium-star nuclei. Their likely progenitors are  wide binaries in which the primary star reached the thermally pulsing asymptotic-giant-branch (AGB) phase, dredged up C and \sprocess\ elements from its interior, and transferred enriched material to the companion through a dense stellar wind. The remnant core is now a hot, optically inconspicuous (pre-)white dwarf, responsible for ionizing the AGB ejecta, and leaving the optical spectrum dominated by the cool barium star. We present deep narrow-band images of \K, obtained by accumulating long exposure times using amateur telescopes. The PN shows a thin-ring morphology, remarkably similar to the ``wedding-ring'' shapes seen around other members of this class of binary PNNi. The thin ring probably represents material preferentially ejected into the orbital plane of the binary; we note that the PNN is slightly off-center within the ring, as has been predicted theoretically. We suggest several follow-up studies, including precision photometry to search for periodic variations due to starspots on the rotating barium star, and high-resolution spectroscopy to determine atmospheric parameters of the star, chemical abundances, and its rotation velocity.

\null\vskip 0.25in

\end{abstract}

%% Keywords should appear after the \end{abstract} command. 
%% See the online documentation for the full list of available subject
%% keywords and the rules for their use.

% \keywords{blah --- blah}

\section{Introduction \label{sec:intro} }

This is the eighth in a series of papers presenting results from a spectroscopic survey of central stars of faint Galactic planetary nebulae (PNe). The survey is carried out using the second-generation ``blue'' Low-Resolution Spectrograph (LRS2-B) of the 10-m Hobby-Eberly Telescope (HET; \citealt{Ramsey1998,Hill2021}), located at McDonald Observatory in west Texas, USA\null. The preparation of our target list of planetary-nebula nuclei (PNNi), an overview of the instrumentation and data-reduction techniques, and some initial discoveries of extremely hot PNNi were presented in our first paper \citep[][hereafter Paper~I]{Bond2023a}. Subsequent publications have described a ``PN mimic'' (i.e., a Str\"omgren zone of the interstellar medium ionized by a passing hot subdwarf, rather than material ejected from it); new extremely hot hydrogen-deficient PNNi; a late-type spotted and rotationally variable central star; and the unusual PNN of Abell~57, which is accompanied by a dust-enshrouded companion star. Our most recent publications [\citealt{Reindl2024}, hereafter Paper~VI; and Werner et al.\ (in press), hereafter Paper~VII] present atmospheric analyses of 17 new hot hydrogen-rich PNNi, increasing the sample of such objects with known astrophysical parameters by about 20\%; and analyses of 30 new hydrogen-deficient central stars. Papers~VI and VII may be consulted for references to our earlier papers.

The spectra of most PNNi are those of extremely hot stars, whose ultraviolet (UV) radiation is responsible for photoionizing their surrounding nebulae. In this paper, however, we present our discovery of a nucleus whose optical spectrum shows a cool solar-type star, one having an unusual chemical composition. It is the central star of Kohoutek $\text{1-9}$ (\K), a faint and relatively little-studied PN, lying in a low-Galactic-latitude field in Monoceros. \K\ is designated PN~G219.3+01.1 in the standard naming convention for PNe, based on Galactic coordinates.

\section{Planetary Nebula K 1-9 and its Nucleus \label{sec:K1-9} }

\subsection{Discovery \label{sec:discovery} }

\K\ was discovered by \citet{Kohoutek1963} in the course of his search for faint PNe on photographic prints of the Palomar Observatory Sky Survey (POSS)\null. He described \K\ as an oval ring with major and minor axes of $46''\times28''$, detected on the red POSS print but not on the blue one.  

%There is a suggestion of an outer envelope with a diameter of about $60''$.

In spite of its morphology, \K's classification as a PN was questioned by \citet{Acker1987} and \citet{Acker1990}. They suggested it is likely to be an \hii\ region instead, on the basis of a low excitation level (weak [\oiii], strong [\nii]) seen in their spectroscopy of the nebula. However, \citet{Kaler1990} obtained spectrophotometry of the central region of the nebula (aperture diameter $8''$) and interpreted \K\ as a low-excitation PN\null. They determined chemical abundances, finding a high N/O abundance ratio. On this basis they classified \K\ as a nitrogen-rich ``Type~I'' PN, as defined by \citet{Peimbert1978}. An analysis of slit-spectroscopic line fluxes by \citet{Kondratyeva2003} led to a minimum effective temperature of the ionizing source of $\sim$54,000~K. 

A narrow-band CCD image of \K\ in \Ha+[\nii] was presented by \citet{Schwarz1992}, clearly showing its oval hollow-ring structure; however, the nebula was only marginally detected in [\oiii] $\lambda$5007. More recently, an image with similar depths in those bandpasses has been posted online\footnote{\url{https://app.astrobin.com/u/GaryI?i=qslr26}} by Gary Imm, based on exposures with his 11-inch telescope.
Further information about \K\ is available in the online Hong-Kong/AAO/Strasbourg/H$\alpha$ Planetary Nebulae (HASH) database\footnote{\url{http://hashpn.space/}} \citep{Parker2016, Bojicic2017}. In the HASH catalog, \K\ is placed in the category of ``true PNe.''

\subsection{Central Star \label{sec:centralstar} }

As noted by \citet{Kohoutek1963}, a faint star lies near the center of \K. Kohoutek did not consider the star to be the nucleus because of its reddish color; he found a blue minus red color index of about +0.7~mag, based on the blue and red POSS prints. However, the marked peculiarities of this star as described below, as well as the lack of an alternative candidate, make it all but certain that it is physically associated with the nebula.
Table~\ref{tab:DR3data} lists basic data for this star, taken from \Gaia\/ Data Release~3\footnote{\url{https://vizier.cds.unistra.fr/viz-bin/VizieR-3?-source=I/355/gaiadr3}} (DR3; \citealt{Gaia2016, Gaia2023}).

\begin{deluxetable}{lc}[h]
\tablecaption{\Gaia\/ DR3 Data for Central Star of \K\ \label{tab:DR3data} }
\tablehead{
\colhead{Parameter}
&\colhead{Value}
}
\decimals
\startdata
RA (J2000) & 07 07 15.605 \\
Dec (J2000) & $-05$ 10 07.48\\
$l$ [deg] & 219.34 \\
$b$  [deg] &  +01.15 \\
Parallax [mas] & $0.604\pm0.071$ \\
$\mu_\alpha$ [mas\,yr$^{-1}$] & $-0.877\pm0.065$ \\
$\mu_\delta$ [mas\,yr$^{-1}$] & $-3.086\pm0.065$ \\
$G$ [mag] &  16.92 \\
$G_{\rm BP}-G_{\rm RP}$ [mag] & $1.06$ \\
\enddata
\end{deluxetable}

A Bayesian analysis of the {\it Gaia\/} astrometric data by \citet{BailerJones2021} yields a distance of $1625^{+130}_{-140}$~pc for the central star. 
The interstellar extinction of the nebula was determined by \citet{Kaler1990} from their measured fluxes of its \Ha\ and \Hb\ emission lines, giving a logarithmic extinction at \Hb\ of $c=0.58$. \citet{Kondratyeva2003} found a similar value of $c=0.60$ from their own spectrophotometry. These values correspond to a reddening of $E(B-V)=0.40$, and an extinction in the \Gaia\/ $G$~band of about 1.03~mag. Assuming that this extinction also applies to the central star, we find an absolute magnitude of $M_G=+4.8\pm0.2$.  This value is consistent with a main-sequence star with a spectral type of about G5~V.\footnote{Based on the tables of intrinsic stellar colors at \url{https://www.pas.rochester.edu/~emamajek/EEM_dwarf_UBVIJHK_colors_Teff.txt} and in \citet{Pecaut2013}.} 

Such a star is incapable of photoionizing the PN, making it likely that it is an optical companion of an unresolved hot UV source---whose presence is indicated by the nebular analysis described in Section~\ref{sec:discovery}. Unfortunately, to our knowledge, there are no UV observations available for direct confirmation of such a companion; for example, the site of \K\ was never imaged by the {\it Galaxy Evolution Explorer\/} (\GALEX; see \citealt{Gomez2023}). 

\section{Hobby-Eberly Telescope Spectroscopy and Classification of the Nucleus}

\subsection{Observations and Reductions}

We included the central star of \K\ in the target list for our HET\slash LRS2-B spectroscopic survey. HET observations are carried out by on-site astronomers in a queue-scheduling mode, as described by \citet{Shetrone2007PASP}.
LRS2-B is an integral-field-unit (IFU) spectrograph. It uses 280 $0\farcs6$-diameter lenslet-coupled fibers covering a $12''\times 6''$ field of view (FOV), which feed two spectrograph units via a dichroic beamsplitter.  The ``UV'' channel of LRS2-B covers the wavelength range 3640--4645~\AA\ at resolving power 1910, while the ``Orange'' channel covers 4635--6950~\AA\ at resolving power 1140. 

Data reduction and absolute calibration are carried out by co-author Zeimann, using the \texttt{Panacea}\footnote{\url{https://github.com/grzeimann/Panacea}} and \texttt{LRS2Multi}\footnote{\url{https://github.com/grzeimann/LRS2Multi}} packages. Full details of the LRS2-B spectrograph and general data-reduction procedures are given by \citet{Chonis2016}, and in Paper~I, respectively.  

For each individual exposure, the spectrum of the target star is extracted using a $1\farcs5$-radius circular aperture. The local background is estimated from an annular region spanning radii of 2.5--$5''$; its spectrum is subtracted from the stellar spectrum, removing both night-sky emission and continuum, and the diffuse nebular emission surrounding the star. The use of IFU spectroscopy allows this background to be modeled locally and contemporaneously within the same exposure, which is particularly advantageous for faint PNe with spatially varying surface brightness.

Individual exposures are combined using inverse-variance weighting. Prior to combination, each exposure is scaled according to its signal-to-noise ratio measured near 5100~\AA, ensuring that frames obtained under better observing conditions contribute preferentially to the final spectrum. The combined spectrum is then resampled to a spacing of 0.7~\AA, as described in previous papers in this series.

Our first spectrogram, in 2024 November, revealed that the central star has an unusual composition. On this basis, and in order to improve the signal-to-noise ratio for this relatively faint and reddish star, we returned \K\ to the scheduling queue for additional observations. Ultimately we accumulated a total of six exposures, some of which suffered from poor observing conditions. 
An observation log for our \hbox{LRS2-B} exposures on \K\ is presented in Table~\ref{tab:exposures}.

%All of the data, with appropriate weighting, were combined into a final spectrum and resampled to a spacing of 0.7~\AA, as described in previous papers in this series.

\begin{deluxetable}{lcc}[h]
\tablecaption{Log of HET LRS2-B Observations of \K\ Central Star \label{tab:exposures} }
\tablehead{
\colhead{Date}
&\colhead{Exposure}
&\colhead{SNR\tablenotemark{a}}\\
\colhead{[YYYY-MM-DD]}
&\colhead{[s]}
&\colhead{}}]
\decimals
\startdata
2024-11-16 & 408 & 63 \\
2025-01-12 & 408 & 33 \\
2025-01-14 & 408 & 48 \\	
2025-01-15 & 407 & 51 \\  
2025-03-18 & 606 & 154 \\
2025-10-27 & 907 & 192 \\
\enddata
\tablenotetext{a}{Signal-to-noise ratio per resolution element in the stellar continuum at 5100~\AA.} 
\end{deluxetable}

%Greg's table of S/N in 50 A interval at 5100 A:
%
%multi_20241116_0000012_exp01_orange.fits 	197.12
%multi_20250112_0000013_exp01_orange.fits* 	104.47
%multi_20250114_0000011_exp01_orange.fits 	151.17
%multi_20250115_0000013_exp01_orange.fits 	159.87
%multi_20250318_0000007_exp01_orange.fits 	480.66
%multi_20251027_0000027_exp01_orange.fits 	600.61
%
%Resolving power is 1140, presumably at the Orange channel central w.l. of 5790, so the resolution is 5.1 A. So the 50 A bandpass has 50/5.1 = 9.8 resolution elements. Assume S/N in above table = sqrt(no. of counts in 50 A window). 

%counts = SN**2
%counts/res element = counts/9.8
%snr per res relement = sqrt(counts/9.8)

%Those are the numbers entered in above table.

\subsection{Classification as a Dwarf Barium Star}

\begin{figure*}
\centering
\includegraphics[width=\textwidth]{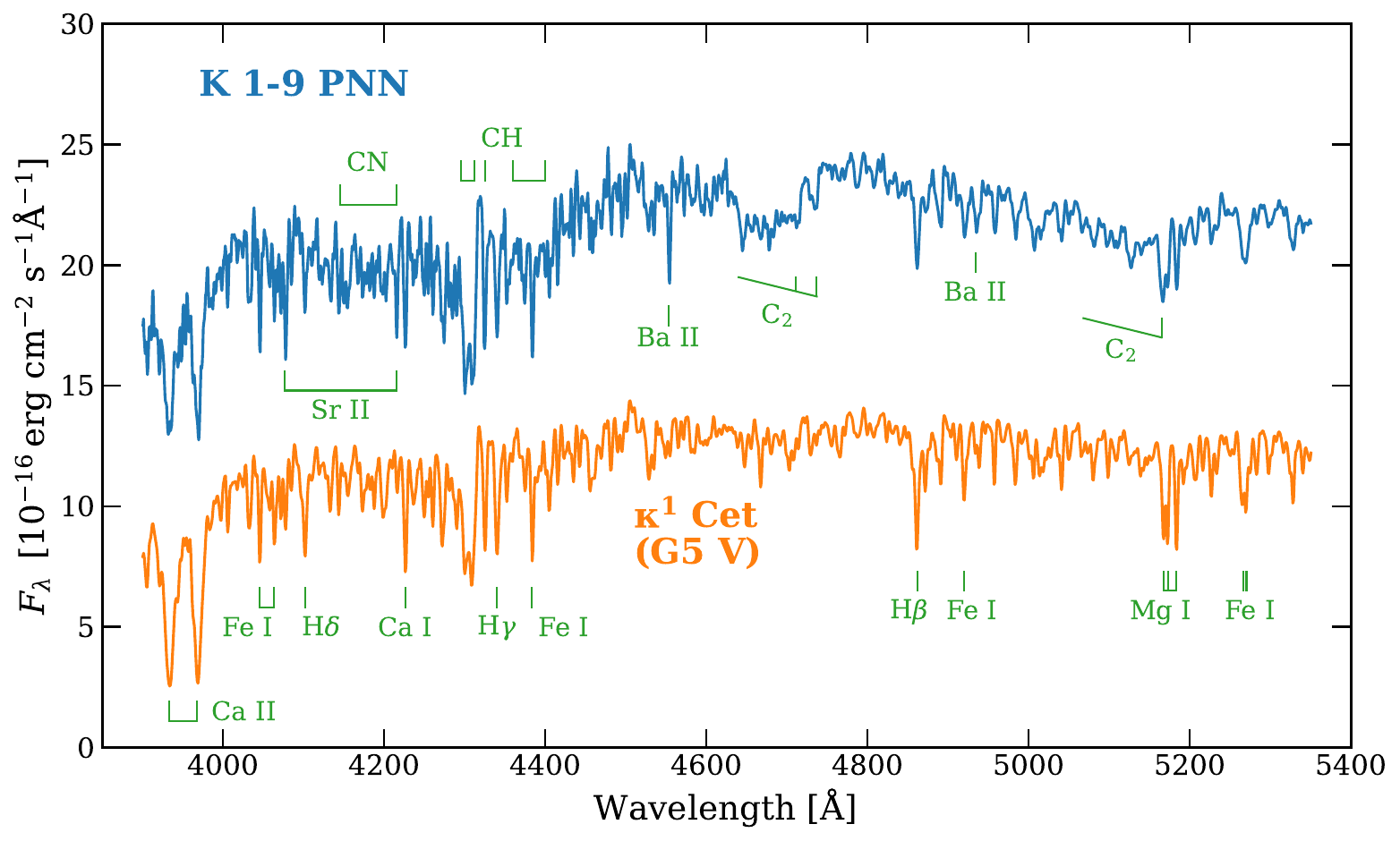}
\caption{
Hobby-Eberly Telescope LRS2-B spectrum of the central star of \K\ (top spectrum, shifted upward by $10^{-15}\rm\,erg\, cm^{-2}\, s^{-1}\,\text{\AA}^{-1}$), and a scaled spectrum of the G5~V standard star $\kappa^1$~Ceti (bottom spectrum). Several atomic lines common to both spectra are labeled in the bottom spectrum, and atomic and molecular features enhanced in \K, due to carbon and \sprocess\ elements, are marked in the top spectrum. 
\label{fig:spectrum}
}
\end{figure*}

Figure~\ref{fig:spectrum} plots our combined LRS2-B spectrum of the \K\ PNN (top spectrum). We have dereddened its absolute fluxes by wavelength-dependent amounts corresponding to $E(B-V)=0.40$; here we adopted the interstellar-extinction formalism of \citet{Cardelli1989} and \citet{ODonnell1994}, and used a standard value of $R_V=A_V/E(B-V)=3.1$. In the figure we have zoomed in on the spectral region from 3900 to 5350~\AA, the usual range used for spectral classification. 

For comparison, the bottom spectrum in Figure~\ref{fig:spectrum} is that of $\kappa^1$~Ceti, a G5~V classification standard.\footnote{See the detailed discussion at \url{https://www.pas.rochester.edu/~emamajek/spt/G5V.txt}} 
This spectrum was downloaded from the MILES website\footnote{The acronym stands for Medium-resolution Isaac Newton Telescope Library of Empirical Spectra. See \url{http://miles.iac.es}} \citep{Falcon2011}, which provides a library of observed spectra of standard stars, with a resolution similar to that of our HET data. The spectrum has been scaled to the absolute flux of \K, and to match the line profiles of the HET spectrum we applied a boxcar smoothing of 3~pixels (2.7~\AA). We then shifted the spectrum of \K\ upward by adding a constant $10^{-15}\rm\,erg\, cm^{-2}\, s^{-1}\,\text{\AA}^{-1}$, in order to separate it from the spectrum of $\kappa^1$~Cet.

Several absorption lines of hydrogen and metals are labeled in Figure~\ref{fig:spectrum} underneath the spectrum of $\kappa^1$~Cet. These features have similar strengths in the \K\ spectrum, consistent with the latter having a spectral type near G5~V, as was expected from its absolute magnitude (Section~\ref{sec:centralstar}). 

However, several absorption features are markedly stronger in \K\ than in the G5~V standard; they are labeled in the top spectrum in the figure. The Swan bands of C$_2$ are conspicuously strong, and the G band of CH and the $\lambda$4215 CN band are likewise enhanced in \K\ relative to the standard star. \K\ also shows extreme enhancements of resonance lines of ionized strontium (\srii\ $\lambda\lambda$4077, 4215) and barium (\baii\ $\lambda$4934 and its {\it raie ultime\/} at $\lambda$4554).  The overabundance of barium, relative to solar, is about $\rm[Ba/Fe]=+1.1\,dex$.\footnote{Based on a follow-up high-resolution spectrum, obtained by S.~Bhattacharjee (priv.\ comm.)}

%extimate the enhancement factors here...

The nucleus of \K\ thus has the properties of a classical barium star (or ``\BaII\
star''). Barium stars were first identified as a spectroscopic class by
\citet{Bidelman1951}.  They are late-type stars showing enhanced abundances of carbon and of heavy elements, including Sr and Ba, produced by slow-timescale neutron-capture (\sprocess) reactions. Following the discovery that essentially all \BaII\ stars are long-period spectroscopic binaries \citep[e.g.,][]{McClure1984}, the modern view is that barium stars are the binary companions of more massive stars that reached the thermally pulsing asymptotic-giant-branch (TPAGB) stage. Carbon and \sprocess\ elements were dredged up from their interiors to their surfaces, and then transferred to the photospheres of the companions via a stellar wind---a process described theoretically by, for example, \citet{Boffin1988}. The AGB components have now become optically invisible (pre-)white dwarfs, leaving the contaminated companions as the visible \BaII\ stars. In order for there to be room for the primary to reach the TPAGB stage, the initial binary must have a sufficiently wide separation and a relatively long orbital period. For reviews of barium stars and their origins, see, for example, \citet{Wallerstein1997, BondSion2000, Izzard2010, Escorza2023review}; and references therein.  

The overall spectral-energy distributions (SEDs) of \K\ and the G5~V standard shown in Figure~\ref{fig:spectrum} are reasonably similar. However, a subtle difference is that the SED of \K\ at wavelengths shortward of $\sim$4600~\AA\ fades more steeply than that of the standard star. This is due to a broadband depression seen in the SEDs of all barium stars, centered around 4000~\AA\ and several hundred angstroms wide, which was discovered by \citet{BondNeff1969}. More recently, the depression has been shown to be due primarily to the enhanced abundance of CH in these stars' atmospheres \citep[][and references therein]{BondC-def2019}.  

We see no clear evidence in our LRS2-B spectrum for the putative hot companion of \K, down to the 3640~\AA\ cutoff of our data. It is difficult to set firm constraints, because of the complex interplay near the blue cutoff between reddening, the enhanced CN 3883~\AA\ band, and the Bond-Neff absorption. Spectroscopy and imaging down to the atmospheric cutoff, and of course from space, would be useful for a direct detection of the companion. 

The \BaII\ stars noted by \citet{Bidelman1951} are G- and K-type red giants. This continued to be the case as more barium stars were discovered throughout the 1950s and 60s \citep[e.g.,][]{Warner1965}. However, the mass-transfer and ``innocent-bystander'' scenario for the creation of these objects should also lead to contamination when the companion is a main-sequence star. This expectation had already been borne out by the discovery \citep{Bond1974} of F- and G-type dwarfs and subgiants with spectral properties similar to those of red-giant \BaII\ stars. A few years later, a related class of dwarf carbon stars was discovered \citep{Dahn1977}. There is now a substantial literature on the compositions and origins of dwarf barium stars and dwarf carbon stars; see, for example, \citet{Green2019, North2020, Zhang2023, Rekhi2026}; and references therein. 

\section{Barium Stars in Planetary Nebulae \label{sec:BaStarsInPNe} }

Our conclusion, based on its spectroscopic properties and relatively low luminosity, is that the nucleus of \K\ is a dwarf barium star.  \K\ thus joins a select group of extremely young barium stars, for which the launching of the AGB wind was so recent that the photoionized ejecta are still visible as a surrounding PN\null.  In this section we briefly review this small class of objects.

%It is a G5~V star whose surface was polluted while it was immersed in a stellar wind from its former AGB primary companion.

\subsection{Abell 35 and LoTr 5}

The nuclei of Abell~35 and LoTr~5 are G-type stars. \citet{Thevenin1997} obtained spectroscopy showing both stars to have modestly enhanced abundances of Ba.\footnote{However, according to S.~Bhattacharjee (priv.\ comm.), a recent high-resolution spectrum of Abell~35 does not confirm an overabundance of barium.} However, they do not exhibit the strong C$_2$ molecular bands seen in \K. Both objects are conspicuous UV sources, indicating the presence of hot companions of the cooler stars that dominate at optical wavelengths.

\citet{Aller2018} showed that the orbital period of the LoTr~5 system is $\sim$2700~days. LoTr~5 is classified as a true PN in HASH\null. To our knowledge, the orbital period of the Abell~35  binary remains unknown, but it is longer than at least several decades \citep[e.g.,][]{Gatti1998}. Although we  refer to Abell~35 as a PNN, its surrounding nebula has a bow-shock morphology. This indicates (as argued by, for example, \citealt{FrewParker2010} and \citealt{Ziegler2012}) that the nebula is likely a PN mimic due to a chance encounter of the hot source and its fast stellar wind with an interstellar cloud, rather than being ionized material ejected from the central binary itself.

\bigbreak

\subsection{WeBo 1}

A cool central star, with pronounced overabundances of carbon and \sprocess\ elements, was discovered in the PN WeBo~1 by \citet[][hereafter BPW03]{Bond2003}. Its optical spectrum (Figure~2 in BPW03) is remarkably similar to that of \K, including the presence of strong C$_2$ Swan bands. However, there are two differences. First, the spectrum of WeBo~1 shown in BPW03 exhibits a strong emission line at \caii\ H $\lambda$3968 (unfortunately the companion \caii\ K line at 3933~\AA\ was outside the wavelength coverage of the spectrogram). BPW03 suggested that the \caii\ emission indicated strong chromospheric activity, due to a rapid rotation of the star. However, we have obtained LRS2-B spectra of WeBo~1 on 10 occasions between 2019 and 2022, and none of them showed emission at \caii\ H and K\null. This appears to indicate that the emission seen by BPW03 was due to a rare and energetic flare event, or conceivably that it was an instrumental artifact of some kind. (We also note that no \caii\ H and K emission was seen in any of our six spectra of \K.)

A second difference is that WeBo~1 appears to be a considerably more luminous star than \K. BPW03 estimated a distance to WeBo~1 of $\sim$1600~pc, using three indirect methods. This implied a visual absolute magnitude for the PNN of $M_V\simeq+1.3$, and a spectral classification of K0~III\null. That result is now updated by the direct trigonometric parallax determined by \Gaia, from which \citet{BailerJones2021} obtain a slightly shorter distance of $1500^{+41}_{-49}$~pc. At this distance, and adopting an extinction of $A_G=1.8$ [based on a reddening of $E(B-V)=0.70$, determined by \citealt{Siegel2012}], we find $M_G=+1.1$. This confirms that the central star of WeBo~1 seen at optical wavelengths is a red giant.

Near-UV observations of WeBo~1 with the UVOT camera onboard the {\it Neil Gehrels Swift Observatory\/} confirmed the presence of a hot companion of the optical \BaII\ star \citep{Siegel2012, Siegel2024}. 

\subsection{Abell 70}

\citet[][hereafter M12]{Miszalski2012} discovered that the nucleus of the PN Abell~70 is a binary containing a G8~IV-V star, along with a hot companion that dominates the spectrum shortward of $\sim$3800~\AA\null. The hot companion is conspicuous in near- and far-UV imaging by \GALEX\null. Based on high-resolution spectroscopy, the cool component shows modest abundance enhancements of Sr and Br, but not of carbon \citep[M12;][]{JonesA702022}.

M12 presented two possible distances for Abell~70, based on statistical relations based on nebular surface brightness, of 2.4 and 5.0~kpc. As it turns out, the \Gaia\/ parallax of the central star ($0.25 \pm 0.12$~mas) implies a similar range of distances of 2.6 to 5.1~kpc \citep{BailerJones2021}. \citet{JonesA702022} estimated a distance of 4.6~kpc, based on a model-atmosphere SED of the cool component and photometry from \Gaia. For this range of distances, the implied absolute magnitude of the cool component of the nucleus ranges from $M_V=+4.2$ to +5.8 (Table~8 in M12). 

% need to mention Jones+ 2022  \citet{JonesA702022}

\subsection{Hen 2-39}

In a program aimed at spectroscopy of PNNi apparently having late spectral types, \citet[][hereafter M13]{MiszalskiHen2-392013} discovered that the central star of the southern-hemisphere PN \Hen\ is a barium star. Its spectrum is similar to that of WeBo~1, but is slightly later, at a type of about K3~III\null. As in the case of WeBo~1 (and \K), the Swan bands of C$_2$ are strongly enhanced, along with \SrII, \BaII, and other \sprocess\ elements. Unfortunately, the \Gaia\/ parallax of the nucleus has an uncertainty of $\sim$45\%. Using indirect estimates of its distance, M13 derived an absolute magnitude of $M_V\simeq+1.6$. This was refined to $M_V\simeq+0.9$ in a detailed spectroscopic analysis by \citet{LoblingHen2-392019}, which confirmed the overabundances of carbon and \sprocess\ elements.

Due to its low Galactic latitude, there is no \GALEX\/ imagery to confirm the presence of a hot companion in Hen 2-29. However, there is little doubt about its existence.

\subsection{The Necklace Nebula}

We mention in passing IPHASX J194359.5+170901 (the ``Necklace'' Nebula; PN\,G054.2$-$03.4). Its central star is a close binary with an orbital period of 1.16~days \citep{Corradi2011,Jones2026}. Spectroscopy of the nucleus by \citet{MiszalskiNecklace2013} revealed the presence of C$_2$ absorption bands, arising in the spectrum of an irradiated cool companion of the hot central star. However, there is no evidence for enhanced abundances of \sprocess\ elements, so the object is not classified as a barium star. \citet{MiszalskiNecklace2013} suggested a scenario in which the initial binary had a separation close enough that the original primary reached the AGB stage and was dredging up carbon. Then, before the onset of thermal pulses and \sprocess\ nucleosynthesis, it filled its Roche lobe and entered into a common-envelope interaction with the companion, followed by a drastic shortening of the orbital period. In contrast, the other objects mentioned in this section are likely to have had much longer initial orbital periods. Their primaries were able to reach the TPAGB stage, allowing them only then to contaminate their distant companions via a stellar wind.

%\bigbreak

%\section{Search for Rotational Light Variability}

\section{\K\ as an Abell 35-Type Central Star}

% \subsection{Variability of Abell~35-type Systems}

Abell~35 itself and LoTr~5, discussed in the previous section, belong to a small group of PNNi for which the name ``Abell 35-type central stars'' was proposed by \citet{Bond1993}. This class is defined as binaries consisting of a rapidly rotating late-type star and an optically faint compact hot companion. \citet{Bond1993} listed the three members of the category known among PNNi at that time: Abell~35, LoTr~1, and LoTr~5. 

A scenario for the origin of Abell~35-type systems was proposed by \citet[][hereafter JS96]{Jeffries1996}: they are created in a situation where an AGB star ejects a dense wind, part of which is captured by a moderately distant companion star. This is of course the same scenario for the origin of barium and dwarf carbon stars outlined in previous sections; but it has the added feature that the companion not only accretes material but also angular momentum from the AGB wind, spinning up its rotation.  JS96 describe the spun-up cool companions as ``wind-accretion induced, rapidly rotating stars (or WIRRing stars).'' Their calculations suggest that the spin-ups can occur in binaries with separations as wide as $\sim$100~AU\null. 

%The same scenario has been discussed for the origin of dwarf carbon stars by, for example, \citet{Green2019}.

%The remnant core of the AGB star has now evolved to be an UV-bright hot white dwarf or pre-white dwarf.

As noted by \citet{Bond1993}, the late-type components of Abell~35, LoTr~1, and LoTr~5 had been found to be periodic photometric variables. 
These variations arise because the rotation of the cool components drives magnetic surface activity. This creates starspots on the stellar surface, and thus photometric variations at the stellar spin period. The periods of the light variations are much shorter than the orbital periods of the binaries, which for most of the systems remain unknown. 

In an earlier paper in the present series \citep[][hereafter Paper~IV]{Bond_Pa27_2024}, we presented our discovery of a new Abell~35-type central star, the nucleus of the PN Pa~27. It has a spectral type of about K0~III, making it very similar to WeBo~1---except that Pa~27 does not show enhanced C$_2$ nor \sprocess\ features. In Figure~4 of Paper~IV, we plotted modern high-precision light curves of the known Abell~35-type variables Abell~35, LoTr~1, and LoTr~5, and of the newly discovered Pa~27, obtained with the {\it Transiting Exoplanet Survey Satellite\/} (\TESS) mission. Paper~IV gives details and references to the original discoveries of the variations of the first three. The spin periods for the four objects are 0.77, 6.40, 5.95, and 7.36~days, respectively. At the high precision of \TESS\/ data, the light curves are remarkably smooth sinusoids. 

Periodic light variations have also been found for the Abell~35-type barium nuclei of WeBo~1, \Hen, and Abell~70. Variability of WeBo~1 was discovered by BPW03, showing a period of 4.69~days. Its sinusoidal light curve from \TESS\/ is included in Figure~4 in Paper~IV\null. The peak-to-peak amplitudes of the variations shown in this figure are small, ranging from about 0.03 (LoTr~5 and WeBo~1) to 0.07~mag (Abell~35 and LoTr~1).   
Unfortunately, the central stars of Abell~70 and Hen 2-39 are too faint and/or crowded for useful \TESS\/ photometry. However, for Abell~70 a period of 2.06~days was discovered in dedicated photometry obtained by \citet{BondCiardulloAbell70_2018}, and confirmed by \citet{JonesA702022}. Similarly, for \Hen\ a sinusoidal variation with a period of 5.46~days was found
in photometric data obtained by M13.

Whether or not the WIRRing star in an Abell~35 system becomes a carbon- and\slash or barium-rich star depends on the evolutionary stage of the AGB primary and the composition of its stellar wind at the onset of its interaction with its companion. An additional factor is the amount of dilution of the accreted material by mixing in the convective envelope of the companion. For example, LoTr~5, as noted above, shows \hbox{\sprocess} enhancements; however, LoTr~1 \citep{Tyndall2013} and Pa~27 (Paper~IV) do not. In the case of a G-type dwarf such as \K, with a shallow convective envelope, it may not take very much accretion to produce substantial overabundances of C and \sprocess\ elements.

We investigated whether photometry of the nucleus of \K\ reveals periodic variations, which would confirm it as a new Abell~35-type central star. Unfortunately, however, due to the star's faintness and the presence of the surrounding nebula and several nearby stars, and the large camera pixels, data from \TESS\/ do not provide useful constraints. Periodogram analyses of data from the ASAS-SN\footnote{\url{https://asas-sn.osu.edu}; \citealt{Shappee2014, Kochanek2017}} and ATLAS\footnote{\url{https://atlas.fallingstar.com/}; \citealt{Tonry2018}} sky surveys likewise were not useful, because of large photometric uncertainties compared to the expected rotational signals; for example, the standard deviation of the $g$~magnitudes obtained from ASAS-SN is about 0.25~mag. We examined data having considerably smaller uncertainties, obtained through forced photometry of images from the Zwicky Transient Facility (ZTF).\footnote{\url{https://irsa.ipac.caltech.edu/Missions/ztf.html}; \citealt{Masci2023}. We thank S.~Bhattacharjee for sending us his forced-photometry reductions of the ZTF data.} These $g$-band magnitudes show a typical standard deviation during each observing season of about 0.028~mag, which is consistent with the formal errors of the individual measurements, and thus implying no significant variability larger than the uncertainties. Unfortunately the time coverage of the ZTF data is sparse, with a cadence averaging generally around 4 to 7~days between individual observations. Lomb-Scargle periodograms of the ZTF data, both the entire set covering several years, and individual seasons, showed no significant periodic variations. Thus, from the available data, we can rule out variability as large as that of Abell~35 and LoTr~1, but periodic variations smaller than those of LoTr~5 and WeBo~1 cannot be excluded. 

%Assuming a high inclination....

%{\bf Do we want to mention or discuss inflation of the stars here, or elsewhere, or simply refer to Bhattacharjee in prep?}

% \subsection{Is the Central Star of K 1-9 Variable?}

%{\bf Discussion of searches for variability and reasons for its absence. Maybe the S/N of the phoeometry just isn't good enough.}

\section{Deep Imaging \label{sec:deepimaging} }

\subsection{Observations}

In order to explore the morphology of the faint nebula \K\ beyond what is available in the literature or from online sky surveys, the amateur co-authors of this paper obtained deep images with four different telescopes. These instruments have apertures ranging from 6 to 24~inches, and are equipped with modern low-noise CMOS cameras. The telescopes are located in Chile, Spain, and Mississippi, USA\null. Exposures were accumulated using filters covering narrow-band \Ha+[\nii]\footnote{The \Ha\ 6562~\AA\ filters used in this work also have some transmission at the neighboring [\nii] emission lines at 6548 and 6583~\AA\null. Given the appreciable strength of these lines in the nebular spectrum \citep[e.g.,][]{Kaler1990, Kondratyeva2003}, [\nii] is contributing significantly.} and [\oiii] $\lambda$5007, and broad-band RGB\null. Table~\ref{tab:imaging_exposures} summarizes the sub-exposure times and numbers of frames. Data were obtained between 2025 March~20 and April~27, and the grand total exposure time was 86.5~hours.

\begin{deluxetable*}{lcccccccc}[h]
\tablecaption{Imaging Exposure Times [s] on 
\K\ \label{tab:imaging_exposures} }
\tablehead{
\colhead{Observer}
&\colhead{Location\tablenotemark{a}}
&\colhead{Telescope}
&\colhead{Camera\tablenotemark{b}}
&\colhead{\Ha+[\nii]}
&\colhead{[\oiii]}
&\colhead{R}
&\colhead{G}
&\colhead{B}
}
\startdata
Goodhew & (1) & Twin APM 6 in & (1) & $35\times300$ & $272\times300$ & $20\times300$ & $20\times300$ & $22\times300$ \\
Stern   & (2) & CDK 17 in	  & (2) & $172\times300$ & $84\times300$ & \dots  & \dots  & \dots  \\
Stern   & (2) & CDK 24 in	  & (2) & $191\times300$ & $58\times300$ &  \dots&  \dots & \dots  \\
Talbot  & (3) & Stellarvue SVX 152T 6 in & (3) & $41\times1200$ & \dots         & \dots & \dots & \dots \\
\noalign{\vskip0.05in}
Total exp.\ [hr] & & & & 46.83 & 34.50 & 1.67 & 1.67 & 1.83 \\
\enddata
\tablenotetext{a}{Telescope locations: (1)~Fregenal de la Sierra, Spain; (2)~Río Hurtado Municipality, Chile; (3)~Stark Bayou Observatory, Ocean Springs,
MS.}
\tablenotetext{b}{CMOS cameras: (1)~QHY 268; (2)~Moravian C5A-100M; (3)~ZWO ASI 6200 MM pro.}
\end{deluxetable*}

Pre-processing of the frames was done using {\tt CCDStack},\footnote{\url{https://ccdware.com/ccdstack_overview}} {\tt PixInsight},\footnote{\url{https://pixinsight.com}} and {\tt Photoshop}.\footnote{\url{https://www.adobe.com/products/photoshop.html}} The large number of frames was combined using {\tt AstroPixelProcessor}.\footnote{\url{https://www.astropixelprocessor.com}}

\subsection{Morphological Features}

\begin{figure*}
\centering
\includegraphics[width=6in]{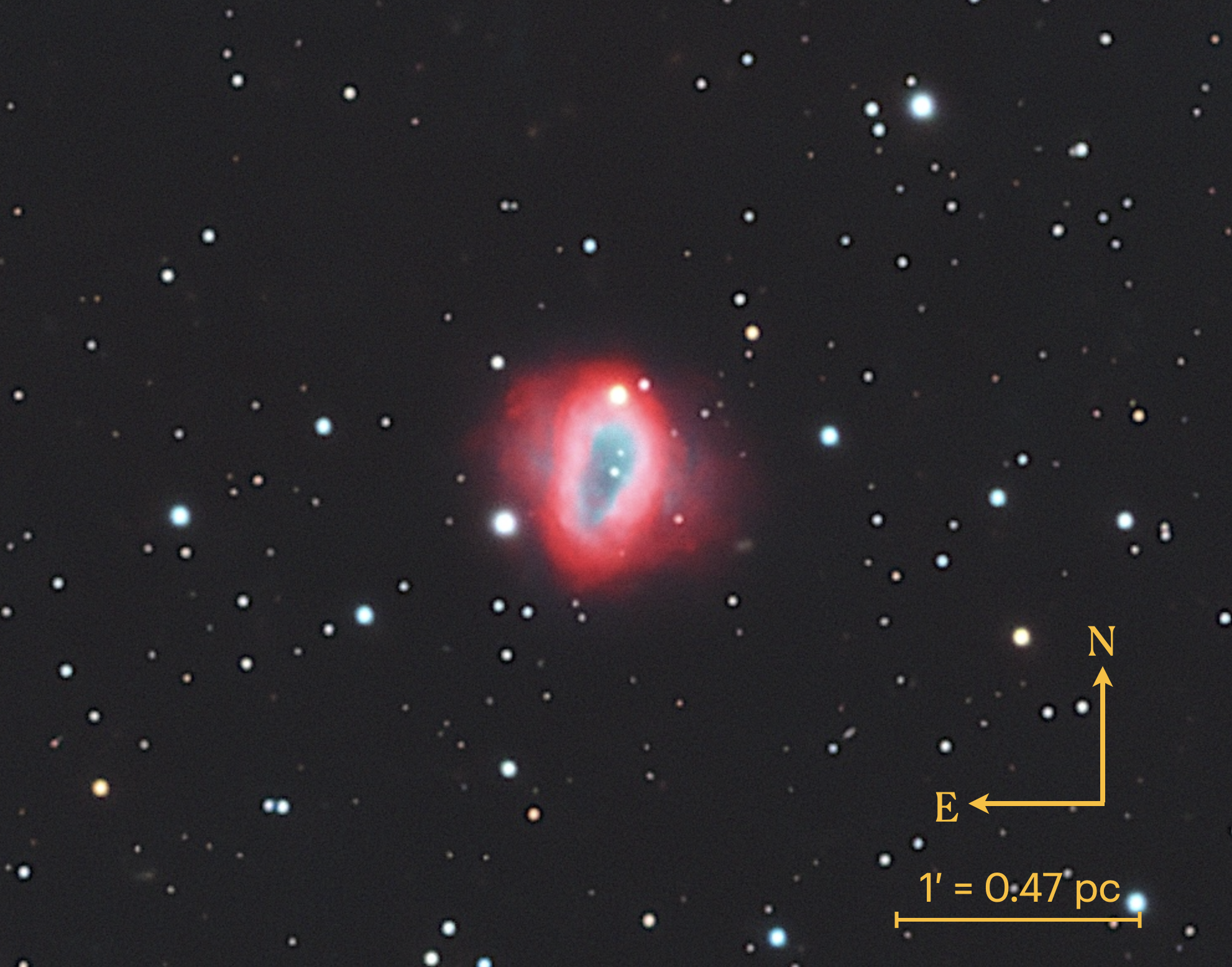}
%\vskip-0.4in
\caption{
Color image of the planetary nebula \K, created from 86.5~hours of exposure time in RGB, \Ha, and [\oiii] $\lambda$5007 filters, as described in the text. \Ha+[\nii] is assigned to the red channel, and [\OIII] $\lambda$5007 to the green and blue channels. Orientation and scale of the image (including a conversion to linear size at the distance of \K) are indicated at the lower right. The carbon- and barium-enriched nucleus lies slightly west of the geometric center of the nebula.
\label{fig:colorimage}
}
%The dwarf barium star nucleus is located slightly west of the center of the nebula.
\end{figure*}

\begin{figure}
\centering
\includegraphics[width=0.47\textwidth]{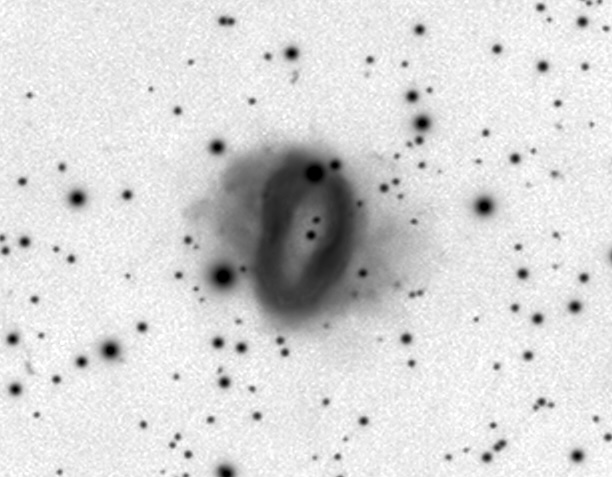}
\includegraphics[width=0.47\textwidth]{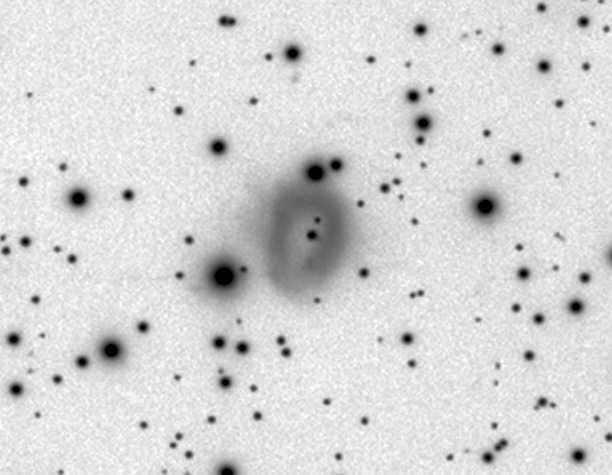}
\includegraphics[width=0.47\textwidth]{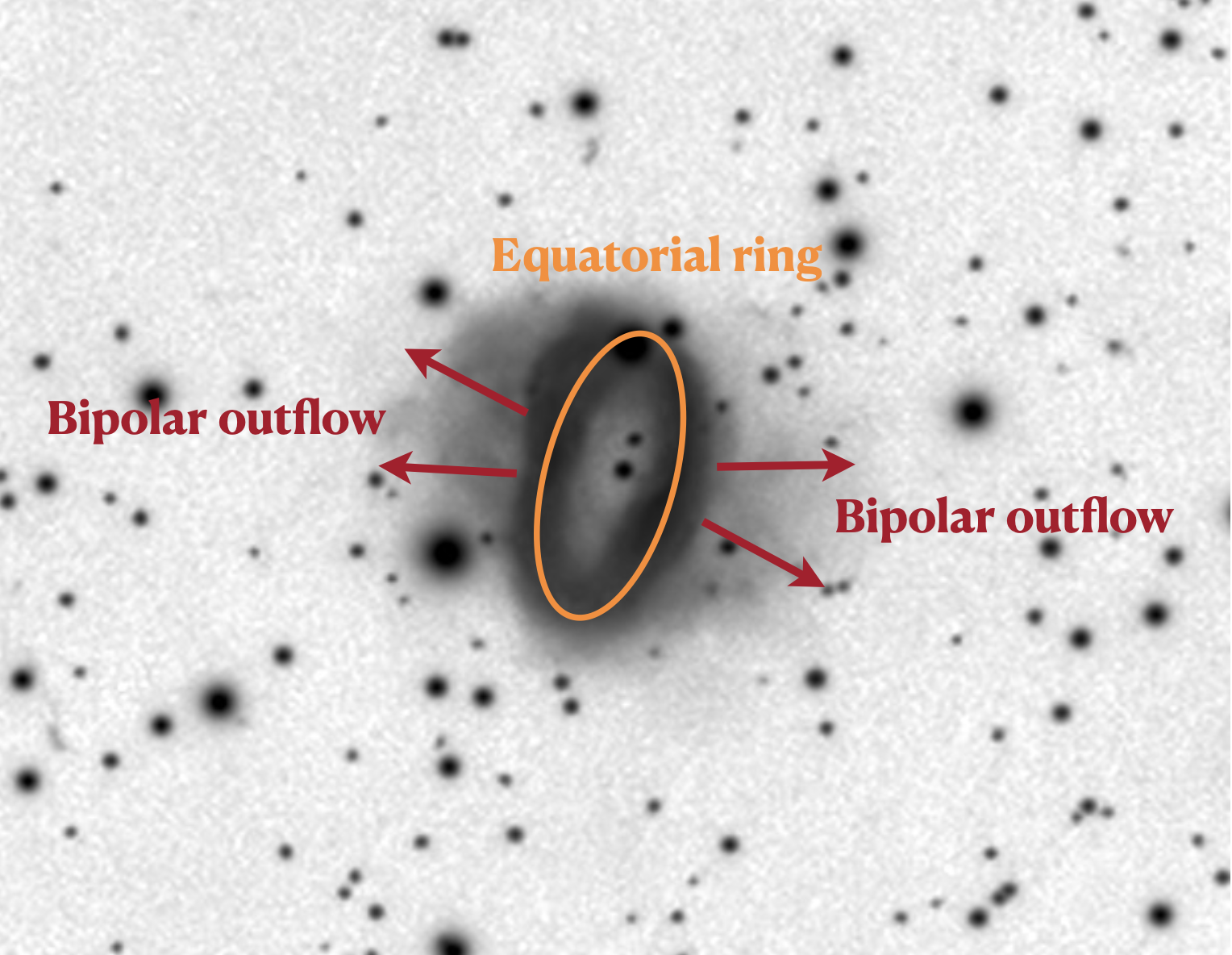}
%\vskip-0.4in
\caption{Narrow-band images of \K\ in \Ha+[\nii] (top) and [\oiii] $\lambda$5007 (middle). In the bottom frame, the equatorial ring and perpendicular bipolar outflows, which are discussed in the text, are labeled on top of the \Ha+[\nii] image. 
\label{fig:BandWframes}
}
\end{figure}

Figure~\ref{fig:colorimage} displays a color picture of the nebula, in a rendition created by combining all of the frames listed in Table~\ref{tab:imaging_exposures}. For this presentation, an ``HOO'' palette was employed, with \Ha+[\nii] assigned to the red channel, and [\OIII] $\lambda$5007 to the green and blue channels. The orientation and angular scale are indicated at the lower right, along with the conversion to a linear scale at the adopted 1625~pc distance of \K. 
Figure~\ref{fig:BandWframes} shows separate frames in \Ha+[\nii] (top) and [\oiii] (middle). 

The images show a nebula with a size of about 0.5~pc. It has a bipolar structure, similar to that often seen in PNe. A bright and slightly distorted hollow elliptical ring extends from the north-northwest to the south-southeast. The excitation level of the nebula is relatively low, as shown especially in Figure~\ref{fig:BandWframes}: the ring is prominent in \Ha+[\nii], but is much fainter in [\oiii]---which explains why it was barely detected in the POSS blue images and the short [\oiii] exposure shown in \citet{Schwarz1992}; see Section~\ref{sec:discovery}. Perpendicular to the bright ring are larger faint bipolar lobes, seen in \Ha+[\nii] but barely visible in [\oiii]\null. The equatorial ring and bipolar lobes are marked schematically on top of the \Ha+[\nii] image in the bottom frame in Figure~\ref{fig:BandWframes}.

The overall structure of \K\  is consistent with a scenario in which an outflow from the AGB component was focused into an equatorial plane due to an interaction with a wide binary companion, producing the bright ring. This process for creating bipolar nebulae from binary central stars has been described theoretically by, for example, \citet{Mastrodemos1999} and \citet{Bermudez2020}, and observationally by \citet{Decin2020}.   Subsequently, when the AGB remnant evolved to a high temperature and developed a fast wind, this wind broke out perpendicularly to the ring, inflating the bipolar lobes.

A comparison of our Figure~\ref{fig:BandWframes} with Figure~1 in BPW03 shows a remarkable similarity of \K\ to WeBo~1. WeBo~1 likewise presents a thin, low-excitation ring, prominent in \Ha+[\nii], and considerably fainter in [\oiii]\null. The inclination angles to our line of sight are even roughly the same. Images of Abell~70 (see M12) likewise show a thin ring, which in this case is only slightly elliptical.

\K\ and WeBo~1 exhibit another similarity: in both objects, the central star is offset from the geometric center of the bright ring, and the displacement is especially conspicuous in \K. Such an offset has been predicted theoretically by \citet{Soker1998}, in the case of AGB wind ejection in an eccentric wide binary, having an orbital period in the range of about 15 to 500~yr.  In Abell~70 and \Hen, however, the stars appear centered within the nebulae (see images in M12 and M13; but it is of course possible that they are displaced along the line of sight). 

The thin- or ``wedding-ring'' morphology seen in \K\ and WeBo~1 is relatively rare among PNe. Four examples are the following: (1)~The nucleus of Hen 2-147 is a symbiotic binary. \citet{Santander2007} presented high-resolution {\it Hubble Space Telescope\/} images of Hen 2-147, revealing a thin ring with an off-center nucleus. Morphologically, it is almost a twin of \K. (2)~SuWt~2 is a thin-ring nebula that also closely resembles WeBo~1 and \K. \citet{Exter2010} discussed a star slightly off-center within the nebula that is a 4.9-day eclipsing binary consisting of two A-type stars. However, on the basis of radial-velocity measurements, \citet{Jones2017} argue that the binary is superposed by chance on the PN----which, if so, makes it a remarkable coincidence. (3)~Sp~1 is a nearly perfect ring, seen face-on \citep{BondLivio1990, Jones2012}. Its central star is a binary with an orbital period of 2.9~days. The optical spectrum is dominated by a hot hydrogen-rich star \citep{Hillwig2016}. (4)~SuWt~3, in the southern hemisphere, is yet another wedding-ring PN \citep[see, e.g.,][]{Schwarz1992}. To our knowledge, its faint central star has not been investigated spectroscopically, but its red color (\Gaia\/ DR3 gives $G_{\rm BP}-G_{\rm RP}=1.74$) indicates a late spectral type and a likely similarity to \K\ and WeBo~1.

\section{Summary and Future Work}

In summary, we have discovered that the central star of the faint PN \K\ has an optical spectrum dominated by a \hbox{G-type} dwarf, exhibiting strong absorption features of molecular carbon and \sprocess\ elements, such as strontium and barium. \K\ thus joins only two other PNNi known to show similar spectra: the nuclei of WeBo~1 and Hen 2-39. As discussed in Section~\ref{sec:BaStarsInPNe}, a few additional late-type PN central stars are known in which carbon, or \sprocess\ elements, but not both, are overabundant.

These rare objects provide us the opportunity to witness, almost in real time, nucleosynthesis of heavy elements in intermediate-mass stars, and the ejection of this material into the interstellar medium for inclusion in subsequent stellar generations and their planets. The barium-enriched central stars appear to arise from wide binary systems, in which the more massive components reached the TPAGB stage, dredged up nuclearly processed material to their surfaces, and developed dense stellar winds. A portion of the wind has accreted onto the companion star. This ``polluted'' star now dominates the optical spectrum, while the hot AGB core has faded into a UV-bright but optically faint \hbox{(pre-)}white dwarf; it is the source of photoionization of the AGB ejecta, producing the surrounding PN.

We present deep \Ha+[\nii] and [\OIII] images of the \K\ nebula, obtained by accumulating long exposure times on amateur telescopes equipped with low-noise detectors. The nebula shows a thin ring surrounding the central binary. As discussed in Section~\ref{sec:deepimaging}, the morphology of \K\ is remarkably similar to that of WeBo~1, another wedding-ring nebula hosting a barium star. Such a shape has been predicted theoretically; it is due to the interaction of the wide binary companion with the ejecta from the AGB primary star, producing an outflow concentrated to the orbital plane. The central stars of both \K\ and WeBo~1 lie slightly off-center within the nebular rings, a phenomenon also predicted from theory.

We suggest several follow-up observations: (1)~UV imaging and\slash or spectroscopy (with \HST, {\it Swift}, or a future mission) would confirm the presence of the putative hot remnant of the former AGB primary star, and possibly constrain its initial mass. (2)~Long-term radial-velocity monitoring of the central star would eventually determine the orbital period of the binary, although the period could be many years or even decades. (3)~An intensive series of precision photometric measurements would establish whether the star exhibits periodic low-amplitude variability, a sign of starspots due to the rotation of the star having been spun up due to accretion from the AGB wind. If so, the central star of \K\ would join the small class of Abell~35-type PNNi, containing rapidly rotating late-type companions of their hot former primaries. (4)~An atmospheric analysis based on high-resolution spectroscopy would establish the effective temperature and surface gravity of the star, allowing a test of the possibility that it is still out of thermal equilibrium (i.e., inflated) due to the recent accretion event. The spectroscopy would also lead to determinations of the star's chemical composition, in particular the abundances of carbon and heavy \sprocess\ elements. It may even be possible to detect the presence of radioactive technetium in this very recently created dwarf barium star. Additionally, the spectra would provide a measurement of the rotational velocity, $v \sin i$, expected to be high if the star has been spun up.

\acknowledgments

We thank the HET queue schedulers and nighttime observers at McDonald Observatory for obtaining the data discussed here.

The Low-Resolution Spectrograph 2 (LRS2) was developed and funded by The University of Texas at Austin McDonald Observatory and Department of Astronomy, and by The Pennsylvania State University. We thank the Leibniz-Institut f\"ur Astrophysik Potsdam (AIP) and the Institut f\"ur Astrophysik G\"ottingen (IAG) for their contributions to the construction of the integral-field units.

We acknowledge the Texas Advanced Computing Center (TACC) at The University of Texas at Austin for providing high-performance computing, visualization, and storage resources that have contributed to the results reported within this paper.

This work has made use of data from the European Space Agency (ESA) mission
{\it Gaia\/} (\url{https://www.cosmos.esa.int/gaia}), processed by the {\it Gaia\/} Data Processing and Analysis Consortium (DPAC,
\url{https://www.cosmos.esa.int/web/gaia/dpac/consortium}). Funding for the DPAC
has been provided by national institutions, in particular the institutions
participating in the {\it Gaia\/} Multilateral Agreement.

Funding for the \TESS\/ mission is provided by NASA's Science Mission directorate.

This research has made use of the SIMBAD database, operated at CDS, Strasbourg, France.

We thank Soumyadeep Bhattacharjee for discussions and communication of results in advance of publication, and John Thorstensen for assistance with ATLAS data.

\bibliography{PNNisurvey_refs}

\end{document}